\documentclass[prb,twocolumn,showpacs,superscriptaddress,floatfix,longbibliography]{revtex4-1}
\usepackage{graphicx,amsfonts,amssymb,amsmath,xspace,dsfont,bbm}
\usepackage[colorlinks=true,citecolor=green,linkcolor=red,urlcolor=blue]{hyperref}

\usepackage{color}
\definecolor{g}{rgb}{.1,0.4,.1} 
\definecolor{b}{rgb}{0,0.2,1}
\definecolor{rouge}{rgb}{0.82,0.,0.}
\definecolor{vert}{rgb}{0.,0.82,0.}
\definecolor{orange}{rgb}{1,0.5,0.}
\definecolor{bleu}{rgb}{0.,0.,0.82}
\definecolor{m}{rgb}{0.82,0.,0.82}
\definecolor{vert2}{rgb}{0.,0.5,0.}
\definecolor{rougeclair}{rgb}{1.0,0.7,0.7}

\newcommand{\id}{\mathbbm{1}}

\newlength{\ldag}
\newcommand{\Scale}[2][4]{\scalebox{#1}{$#2$}}%

\newcommand{\SU}{{{SU}(2)_2}}

\begin{document}

\title{Ising versus $\SU$ string-net ladder}

\author{Julien Vidal}
\email{vidal@lptmc.jussieu.fr}
\affiliation{Sorbonne Universit\'e, CNRS, Laboratoire de Physique Th\'eorique de la Mati\`ere Condens\'ee,
LPTMC, F-75005 Paris, France}

\begin{abstract}

We consider the string-net model obtained from $\SU$ fusion rules. These fusion rules are shared by two different sets of anyon theories.  In this work, we study the competition between the two corresponding non-Abelian quantum phases in the ladder geometry. A detailed symmetry analysis shows that the nontrivial low-energy sector corresponds to the transverse-field cluster model that displays a critical point described by the $so(2)_1$ conformal field theory. Other sectors are obtained by freezing spins in this model. 

\end{abstract}

\maketitle

%
%
\section{Introduction}
%
%
Since Leinaas, Myrheim, and Wilczek's pioneering works~\cite{Leinaas77,Wilczek82}, anyons have been the subject of intense research. These last two decades, these exotic quasiparticles have triggered much attention because of their potential use for  quantum computers \cite{Nayak08}. In two dimensions, their nontrivial braiding properties may indeed be a key ingredient to perform operations, and their robustness with respect to perturbations provides an efficient protection against decoherence \cite{Kitaev03}. Anyons are also inextricably linked to topologically ordered phases of matter and long-range entanglement (see Ref.~\cite{Wen17} for a recent review). Although a complete classification of these phases is still lacking, substantial progress has been recently achieved for bosonic topological orders \cite{Schoutens16,Wen16}. 

Some of these phases naturally emerge in microscopic models. In particular, the string-net model proposed by Levin and Wen in 2005 \cite{Levin05} allows for the realization of any doubled achiral topological phases and provides an ideal framework to study phase transitions that may arise in the presence of sufficiently strong external perturbations~ \cite{Gils09_1,Gils09_3,Burnell12,Schulz13,Schulz14,Schulz15,Dusuel15,Schulz16_1,Schulz16_2,Mariens17}. In the string-net model, most perturbations considered so far are responsible for a transition between a topological phase and a trivial (non topological) phase.  These transitions are often successfully described by the theory of anyon condensation, also known as topological symmetry breaking, which is the counterpart of the Landau symmetry-breaking theory for conventional phases~\cite{Bais09_1,Burnell11_2,Mansson13,Kong14,Eliens14,Neupert16,Burnell18}.
 
The goal of the present paper is to study phase transitions between two nontrivial phases. To this aim, we consider the string-net model in the ladder geometry for microscopic degrees of freedom obeying $\SU$ fusion rules given in Eqs.~(\ref{eq:fusion1}) and (\ref{eq:fusion2}). This system is likely the simplest where the competition between two non-Abelian theories can be considered on equal footing. There are two families of anyon theories obeying $\SU$ fusion rules, giving rise to two distinct quantum string-net condensed phases. These non-Abelian theories can be distinguished by their Frobenius-Schur indicator, which encodes their behavior with respect to bending operations \cite{Kitaev06,Rowell09}. 
The main result of this work is that the low-energy effective Hamiltonian corresponds to a transverse-field cluster model, which is known to display a second-order quantum phase transition described by the $so(2)_1$ conformal field theory \cite{Lahtinen15_1}. Higher-energy sectors are obtained by freezing spins in this model.

%
%
\section{Fusion rules and Hilbert space}
%
%
Following Levin and Wen \cite{Levin05}, we consider degrees of freedom defined on the links of a trivalent graph. In the present case, these links can be in three different states: $1, \sigma$, and $\psi$. The Hilbert space is spanned by the set of configurations satisfying the branching rules, which stem from the $\SU$ fusion rules 
%
%
\begin{eqnarray}
1 \times a = a \times 1&=& a, \:\: \forall a \in \{1,\sigma,\psi\}, \label{eq:fusion1} \\
\sigma \times \sigma= 1+\psi, \:\: \sigma \times \psi&=& \psi \times \sigma=\sigma, \:\: \psi\times \psi= 1.
\label{eq:fusion2}
\end{eqnarray} 
%
%
These rules imply, for instance, that if two links of a given vertex are in the state $\sigma$, the third link must be in the state $1$ or $\psi$. Violations of these constraints lead to states that are not considered here (charge excitations). For any trivalent graph with $N_{\rm v}$  vertices, the dimension of this Hilbert space is given by \cite{Gils09_3}
%
%
\begin{equation}
\label{eq:dimH}
\dim \mathcal{H}= 2^{N_{\mathrm v}+1}+2^{N_{\mathrm v}/2}.
\end{equation} 
%
%

There are 16 anyon theories obeying the aforementioned fusion rules. These theories can be divided into two sets according to the Frobenius-Schur indicator of  $\sigma$ that can take two different values $\varkappa_\sigma=\pm 1$. Each set contains eight theories that have the same $F$-symbols, but different $S$-matrix and $T$-matrix \cite{Kitaev06,Rowell09}. 

%
%
\section{The string-net model}
%
%
According to the string-net construction \cite{Levin05}, we can build, for any input theory, operators that project onto any state of the corresponding doubled (achiral) theory. 
In their seminal paper \cite{Levin05}, Levin and Wen detailed the action of these operators in terms of the $F$-symbols of the theory considered. This procedure is valid for theories with positive Frobenius-Schur indicators but one must be careful when a string $s$ has a negative $\varkappa_s$. As will be shown in a forthcoming paper~\cite{Simon18}, a simple way to properly take into account such a situation is to replace the quantum dimension $d_s>0$  of the particle $s$ by $\varkappa_s d_s$. Note that this prescription reproduces the result for the semion \mbox{theory \cite{Levin05}}, derived by considering a negative quantum dimension.

Although Ref.~\cite{Levin05} focuses on hexagonal plaquettes, it is straightforward to obtain the action of these projectors on any type of plaquette (see  Appendix \ref{app:Bp}). Matrix elements of these operators only involve $F$-symbols and $\varkappa_s$. Consequently, for the problem at hand, these projectors are identical for each set of theories with the same $\varkappa_\sigma$. For convenience, in the following, we will alternatively refer to Ising theory for the set where $\varkappa_\sigma=+1$ and to $\SU$ for the set where $\varkappa_\sigma=-1$. \\

To study the competition between DIsing and D$\SU$ topological phases (prefix D stands for doubled and achiral \cite{Levin05}), we consider the Hamiltonian
%
%
\begin{equation}
H= - \cos \theta\: \sum_p B_p^{1^+} -\sin \theta \: \sum_p B_p^{1^-},
\label{eq:ham_LW}
\end{equation} 
%
%
where $p$ runs over all plaquettes of the system. Operators $B_p^{1^+}$ and $B^{1^-}_{p}$ are projectors onto the vacua ${1^+}$ and ${1^-}$ of DIsing and D$\SU$ theories in the plaquette $p$, respectively. We refer the reader to Refs.~\cite{Burnell12,Schulz16_1} for a discussion of these doubled phases and their particle content. Importantly, these operators mutually commute except when $(p,p')$ correspond to neighboring plaquettes where $[B_p^{1^+},B^{1^-}_{p'}] \neq 0$. Furthermore, when acting on the same plaquette $p$, they are related by the following identity:
%
%
\begin{equation}
B_p^{1^+}= (-1)^{N_{l_\sigma}} B^{1^-}_{p} (-1)^{N_{l_\sigma}},
\label{eq:unit}
\end{equation} 
%
%
where $N_{l_\sigma}$ is the operator that counts the total number of loops made of $\sigma$ links. Hence, $B_p^{1^+}$ and $B_p^{1^-}$ are unitarily equivalent and the spectrum $H$ is invariant under the transformation $\theta \leftrightarrow \pi/2-\theta$. 

Interestingly, for any $\theta$, the Hamiltonian commutes with $B_p^{\sigma^+}$ and $B^{\sigma^-}_{p}$ that are projectors onto the states ${\sigma^+}$ and ${\sigma^-}$ of DIsing and D$\SU$ theories in the \mbox{plaquette} $p$. By construction, one indeed has $[B_p^{1^\pm},B_{p'}^{\sigma^\pm}]=0$ for all $(p,p')$ but, as shown in Appendix \ref{app:Bp}, one further has here $B_p^{\sigma^+}=B_p^{\sigma^-}$.

Finally, depending on the system topology, one may also have other (nonlocal)  conserved quantities measuring the presence of $\sigma^\pm$ in noncontractible loops as we shall now see in a concrete example.  


%

%
%
\section{The two-leg ladder and Hamiltonian symmetries}
\begin{figure}[t]
\includegraphics[width=0.4\columnwidth]{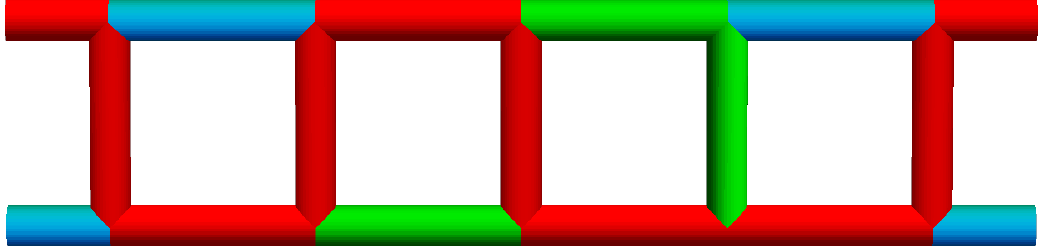} \hfill
\includegraphics[width=0.4\columnwidth]{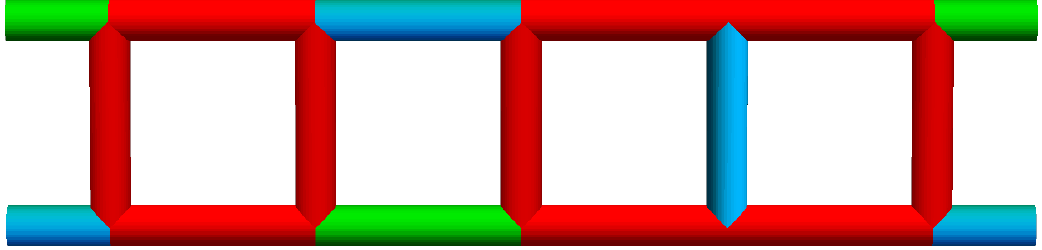}
\caption{Pictorial representation of states belonging to the odd (left) and  even (right) sectors. Blue, green, and red links represent $1,\psi$, and $\sigma$ states, respectively.}
\label{fig:odd_even}
\end{figure}

%
%
In the present work, we focus on a two-leg ladder with periodic boundary conditions. In this geometry, fusion rules given in Eqs.~(\ref{eq:fusion1}) and (\ref{eq:fusion2}) imply that the Hilbert space of this system decouples in two different sectors \cite{Gils09_3}. Indeed, strings of $\sigma$ can only form closed loops since $\sigma \times 1=\sigma \times \psi=\sigma$  and $\sigma \notin \sigma \times \sigma$. In the so-called odd sector, each plaquette has only one leg with a $\sigma$ link and there is only one loop of $\sigma$ links going around the ladder. Hence, in this sector, $B_p^{1^+}$ and $B_p^{1^-}$ have the same matrix elements [see Eq.~(\ref{eq:unit})]. By contrast, in the even sector, each plaquette has either no leg or two legs with a $\sigma$ link and there can be closed loops encircling plaquettes (see Fig.~\ref{fig:odd_even}). Let us note that a similar decoupling exists for $\mathbb{Z}_2$ fusion rules \cite{Morampudi14}.

%
%
%
%

As argued in Refs.~\cite{Gils09_1,Schulz15}, the string-net Hamiltonian defined on a two-leg ladder with periodic boundary conditions also preserves the flux above and below the ladder. In the present case, for any $\theta$, $H$ only commutes with $P_a^{\sigma^\pm}$ and $P_b^{\sigma^\pm}$ that are projectors onto the flux $\sigma^\pm$ above and below the ladder, respectively. As shown in Appendix \ref{app:Bp}, one has  $P_{a,b}^{\sigma^+}=P_{a,b}^{\sigma^-}$ so that we will omit superscript $\pm$ in the following (idem for $B_p^{\sigma^\pm}$).

Projectors $P_a^{\sigma}$, $P_b^{\sigma}$, and $B_p^{\sigma}$ only involve loops of $1$ and $\psi$ ($S_{\sigma \sigma}=0$). As a direct consequence, $\sigma$ links are left unchanged by these projectors since \mbox{$\sigma \times 1=\sigma \times \psi=\sigma$}. Thus, these mutually commuting operators preserve the decoupling between odd and even sectors so that, {\it in fine}, the Hamiltonian can be decomposed as follows:
%
%
\begin{eqnarray}
H&=&H_{\rm odd} \oplus H_{\rm even},\\
&=&\underset{p_a^\sigma,p_b^\sigma,\{b_p^\sigma\}}{\oplus} H_{\rm odd}^{(p_a^\sigma,p_b^\sigma,\{b_p^\sigma\})} \oplus H_{\rm even}^{(p_a^\sigma,p_b^\sigma,\{b_p^\sigma\})},
\end{eqnarray} 
%
%
where $p_a^\sigma,p_b^\sigma$, and $b_p^\sigma$ are the eigenvalues of $P_a^{\sigma}$, $P_b^{\sigma}$, $B_p^{\sigma}$, respectively. Let us mention that fusion rules impose that if $N_\sigma=\sum_p b_p^\sigma$ is even (odd), then $p_a=p_b$ ($p_a \neq p_b$). 

In each sub-sector indexed by  $(p_a^\sigma,p_b^\sigma,\{b_p^\sigma\})$, one can easily generate an eigenbasis by considering states
%
%
\begin{eqnarray}
\label{eq:state}
|p_a^\sigma,p_b^\sigma,\{b_p^\sigma\},  \phi \rangle = &{\mathcal N} & 
\big[p_a^\sigma P_a^{\sigma} +(1- p_a^\sigma) (\mathbbm{1}-P_a^{\sigma})\big]  \nonumber \\
&\times&\big[p_b^\sigma P_b^{\sigma} + (1- p_b^\sigma) (\mathbbm{1}-P_b^{\sigma})\big]  \nonumber \\
&\times& \underset{p}{\Scale[1.5] \Pi}\big[b_p^\sigma B_b^{\sigma} \hspace{0.5pt} + \hspace{0.5pt} (1-b_b^\sigma) (\mathbbm{1}-B_b^{\sigma})\big]  | \phi  \rangle, \qquad
 \end{eqnarray} 
%
%
where $| \phi \rangle$ is a reference state and ${\mathcal N}$ is a normalization factor. Of course, $|\phi \rangle$ may be annihilated by the action of these operators and different reference states may lead to the same final state so that one has to carefully check the completeness of this basis. 

\begin{figure}[h]
\includegraphics[width=\columnwidth]{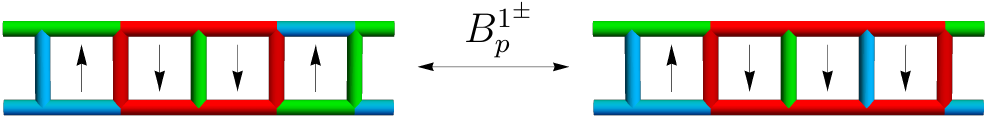}
\caption{Pictorial representation of two even reference states and their spin representation. In the sub-sector (0,0,0), these reference states lead to different states connected by the operator $B_{p}^{1^\pm}$ acting on the rightmost plaquette. }
\label{fig:flip_flop}
\end{figure}

Operators $B_p^{1^+}$ and $B_p^{1^-}$ appearing in the Hamiltonian involve loops of  $\sigma$ ($S_{1 \sigma} \neq 0$) that provide dynamics to the $\sigma$ links.  Roughly speaking, these operators shrink or extend $\sigma$ loops as schematically illustrated in Fig.~\ref{fig:flip_flop}. 
%
%
%
%
This property further decouples each sub-sector indexed by $(p_a^\sigma,p_b^\sigma,\{b_p^\sigma\})$ into sub-sub-sectors. Indeed, in the even sector, if a state has a plaquette $p$ with $b_p^\sigma=1$ and either two or no $\sigma$ legs, it conserves this property under the action of the Hamiltonian. 
Furthermore, the number of plaquettes $b_p^\sigma=1$ with (upper and lower) $\sigma$ legs  must be even, otherwise $\mathcal N=0$. For instance, in the sub-sector $\mbox{$(0,0, b_{p_1}^\sigma=b_{p_2}^\sigma=1)$}$ where only two plaquettes $p_1$ and $p_2$ contain a $\sigma^\pm$ flux, there are two sub-sub-sectors spanned by reference states with or without $\sigma$ loops encircling $p_1$ and $p_2$, respectively. 

This result is easily generalized for any value of \mbox{$N_\sigma=\sum_p b_p^\sigma$}. For a given sub-sector $(p_a^\sigma,p_b^\sigma,\{b_p^\sigma\})$ corresponding to a given $N_\sigma \geqslant 1$, one has $2^{N_{\sigma}-1}$ distinct sub-sub-sectors of dimension $2^{N_{\rm p}-N_{\sigma}}$ ($N_{\rm p}$ being the total number of plaquettes). 
For $(p_a,p_b)=(0,0)$ and $N_\sigma=0$, there is only one sub-sub-sector of dimension $2^{N_{\rm p}}$.

To summarize,  $H$ can be splitted into two sectors, odd and even, according to the parity of $\sigma$ links on the legs of the ladder. In each sector, one can further block-diagonalize $H$ in different sub-sectors according to the presence of a $\sigma^\pm$ flux in plaquettes (measured by \mbox{$B_p^{\sigma}$}) as well as above and below the ladder (measured by $P_a^{\sigma}$ and $P_b^{\sigma}$). Each of these sub-sectors then splits into sub-sub-sectors according to the position of $\sigma$ loops in the reference states. 

Keeping in mind that the dimension of the odd sector is $4^{N_{\rm p}}$ \cite{Gils09_3}, the aforementioned considerations allow for the following decomposition of  the Hilbert space dimension: 
%
%
\begin{eqnarray}
\dim \mathcal{H}&=&\dim \mathcal{H}_{\rm odd}+\dim \mathcal{H}_{\rm even},  \\
&=& 4^{N_{\rm p}}+ 2 
\Bigg[
2^{N_{\rm p}}+ \sum_{N_{\sigma}=1}^{N_{\rm p}} 
\left(
\begin{array}{c}
N_{\rm p}
\\
N_{\sigma}
\end{array}
\right) 
2^{N_{\rm p}-N_{\sigma}} 2^{N_{\sigma}-1} 
\Bigg], \nonumber
\end{eqnarray} 
%
%
where the binomial coefficient simply arises from the different ways to choose the $N_{\sigma}$ plaquettes carrying a $\sigma^{\pm}$ flux among $N_{\mathrm p}$. The factor of 2 in front of the bracket comes from the fact that, in the even sector, the spectrum of $H$ is the same for $(p_a,p_b)=(0,0)$ and $(1,1)$, as well as for $(p_a,p_b)=(0,1)$ and $(1,0)$.
With periodic boundary conditions, one has $N_{\rm p}=N_{\rm v}/2$, so that one directly recovers Eq.~(\ref{eq:dimH}).

%
%
\section{Low-energy sectors}
%
%
To discuss the phase diagram, the first step consists of identifying the sector(s) in which the ground state lies. To achieve this goal, we performed exact numerical diagonalizations of the Hamiltonian. The location of the ground state as a function of $\theta$ is given in Table \ref{tab:main}. 

For $\theta \in ]\pi,3\pi/2[$, ground states are found in all sectors where each plaquette with $b_p^\sigma =0$ are surrounded by two plaquettes with $b_p^\sigma = 1$. Since one has $[B_p^{1^+},B^{1^-}_{p}] = 0$, eigenstates of $H$ in these sectors are simultaneous eigenstates of $B_p^{1^+}$ and $B_p^{1^-}$ with eigenvalues $b_p^{1^+}$ and $b_p^{1^-}$, respectively. The corresponding energy is 
%
\begin{equation}
E(\{b_p^{1^+}\},\{b_p^{1^-}\})=-\cos \theta \sum_p b_p^{1^+} - \sin \theta \sum_p b_p^{1^-}.
\end{equation} 
%
%
Hence, in this parameter range, the ground-state energy per plaquette is $e_0=0$. For all other values of $\theta$, the ground state is always obtained found in a sector where $N_\sigma=0$.

%
%
\begin{table}[h]

\begin{tabular}{c c c c c }
\hline
\hline
$\theta$ & parity & $N_\sigma$ & $(p_a,p_b)$ & degeneracy\\ 
\hline
$]0,\pi/2[$   & odd   & 0 & $(0,0)$ & $1$ \\  
$]\pi/2,\pi[$ & even & 0 & $(0,0),(1,1)$ & $1+1$ \\  
$]\pi,3\pi/2[$ & -- & -- & -- & -- \\  
$]3\pi/2,2\pi[$ & even & 0 & $(0,0),(1,1)$ & $1+1$\\
\hline
\hline
\end{tabular}
\caption{Ground-state sector as a function of $\theta$. }
\label{tab:main}
\end{table}
%
%

For $\theta=0$, the system is, by construction, in a DIsing phase. The degeneracy of $k$th energy level $E_k=-N_{\rm p}+k$ is 
%
%
\begin{equation}
\label{eq:deg}
\mathcal{D}_k=\left(
\begin{array}{c}
N_{\mathrm p}
\\
k
\end{array}
\right)\times 
\left(1+2\times 3^k\right), 
\end{equation} 
%
%
where the binomial coefficient stems from the different ways to choose $k$ plaquettes among $N_{\mathrm p}$ carrying the nontrivial flux excitations. One can check that $\displaystyle{\dim \mathcal{H}=\sum_{k=0}^{N_{\rm p}} \mathcal{D}_k}$. Importantly, one expects $\mathcal{D}_0=3$ ground states with $e_0=E_0/N_{\mathrm p}=-1$: two of them are found in the even sector and the third one lies in the odd sector. The same results hold for $\theta=\pi/2$ where the system is in a D$\SU$ phase.


%
%
\section{Spectrum of the odd sector}
%
%
As already mentioned, in the odd sector, $B_p^{1^+}$ and $B_p^{1^-}$ have the same matrix elements so that each energy level in $H_{\rm odd}^{(p_a^\sigma,p_b^\sigma,\{b_p^\sigma\})}$ is indexed by an integer that simply counts the number of plaquettes carrying a trivial flux. More precisely, one has  
%
\begin{equation}
E(\{b_p^{1}\})=-(\cos \theta+ \sin \theta) \sum_p b_p^{1},
\end{equation} 
%
%
where $b_p^{1}=0$ or $1$ is the eigenvalue of $B_p^{1^\pm}$ in the corresponding eigenstate. Trivial transitions stemming from level crossings belonging to different sub-sectors are observed for \mbox{$\theta=3\pi/4,7\pi/4$}.

%
%
\section{Spectrum of the even sector}
%
%
The nontrivial part of the spectrum is found in the even sector where the Hamiltonian can be written in a simple form. Indeed, in each sub-sub-sector determined by the quantum numbers $(p_a^\sigma,p_b^\sigma,\{b_p^\sigma\})$ and the position of the $\sigma$ loops, the state of each plaquette with $b_p^\sigma=0$ is described by a $\mathbb{Z}_2$ variable that can be interpreted as the flux inside this plaquette (for instance, $1^{+}$ or $\psi^{+}$). A simple way to encode a state $|p_a^\sigma,p_b^\sigma,\{b_p^\sigma\}\rangle$ is to associate an effective spin-$1/2$ configuration to its reference state $|\phi\rangle$. By convention, for any link configuration, we will say that a plaquette is in the state $|\! \! \uparrow \rangle$ if it has no $\sigma$ legs and in the state $|\! \! \downarrow \rangle$ if it has two $\sigma$ legs (see Fig.~\ref{fig:flip_flop} for illustration). In this spin representation,  $B_p^{1^+}$ acts effectively as \mbox{$(\id +\sigma_p^x)/2$} on the spin variable located on plaquette $p$ whereas $B_p^{1^-}$ acts as \mbox{$(\id -\sigma_{p-1}^z \sigma_p^x \sigma_{p+1}^z)/2$}. One can check that these expressions of $B_p^{1^\pm}$ are compatible with Eq.~(\ref{eq:unit}).
In addition, as already mentioned, the presence (or the absence) of $\sigma$ legs in plaquettes with $b_p^\sigma=1$ is preserved by the Hamiltonian. In the spin language, it means  that spins located in plaquettes with $b_p^\sigma=1$ are frozen and $B_p^{1^\pm}$ acting on these plaquettes gives 0.

%
%
\subsection{$N_\sigma=0$}
%
%
In the sub-sectors without $\sigma^\pm$ flux, all spins can freely fluctuate so that the Hamiltonian reads
%
%
\begin{equation}
H_{\rm even}^{(0,0,0)}=-\cos \theta \sum _p \frac{\id + \sigma_p^x}{2} - \sin \theta \sum_p \frac{\id -\sigma_{p-1}^z \sigma_p^x \sigma_{p+1}^z}{2}.
\end{equation} 
%
%
This model, known as the transverse-field cluster model (TFCM) \cite{Pachos04} can be solved exactly using the standard Jordan-Wigner transformation \cite{Montes12,Lahtinen15_1}. In the thermodynamical limit, the ground-state energy per plaquette can be written as
%
%
\begin{equation}
e_0 = -\frac{1}{2}\bigg[ \cos\theta+\sin\theta+\frac{2}{\pi} \left( \frac{2+g}{g} \right)^{1/2} {E}\left(\frac{4}{2+g}\right) \bigg], \qquad \qquad 
\end{equation} 
%
%
where $\displaystyle{E(k)= \int_0^{\frac{\pi}{2}} {\mathrm d} \theta\sqrt{1-k^2 \sin^2 \theta} }$ is the complete elliptic integral of the second kind with \mbox{$g=\tan \theta+(\tan \theta)^{-1}$}, and the low-energy gap is
%
%
\begin{equation}
\Delta = \min\left(|\cos\theta+\sin\theta|,|\cos\theta-\sin\theta| \right). 
\end{equation} 
%
%
Hence,  the phase diagram in this sub-sector $(0,0,0)$ consists of four equivalent gapped phases \cite{Montes12,Verresen17} with a unique ground state separated by four transition points at $|\tan \theta|=1$. As explained in Ref.~\cite{Lahtinen15_1}, with periodic boundary conditions, these critical points are described by the $so(2)_1$ conformal field theory.

%
%
\begin{figure}[h]
\includegraphics[width=\columnwidth]{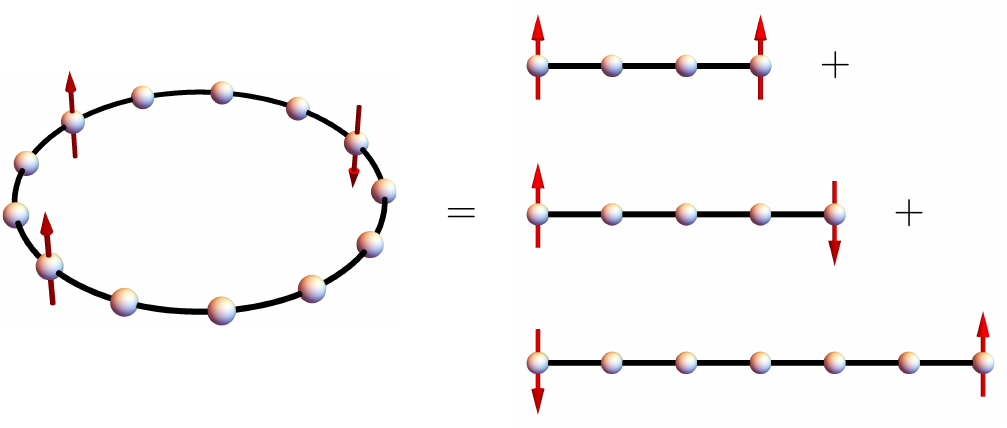}
\caption{Sketch of the decoupling induced by frozen spins (red) in the TFCM. The sub-sub-sector displayed here corresponds to $N_\sigma=3$ where two plaquettes with $b_p^\sigma=1$ have two $\sigma$ legs and the third one has no  $\sigma$ legs. }
\label{fig:decoupling}
\end{figure}
%
%

%
%
\subsection{$N_\sigma \geqslant 1$}
%
%
For sub-sectors where $N_\sigma \neq 0$, the situation is slightly different since one must then consider the TFCM with frozen spins. Actually, these frozen spins effectively cut the system into several subsystems as shown in Fig~\ref{fig:decoupling}. As a result, it is sufficient to study the spectrum of the TFCM with frozen spins at the boundaries \cite{Configuration}  to determine the spectrum of all sectors with $N_\sigma \geqslant 1$. The influence of the boundary conditions manifests notably in the finite-size critical spectrum \cite{Cardy89} but phase transitions that may occur in the thermodynamical limit in sectors with $N_\sigma \geqslant 1$ are the same as in the \mbox{$N_\sigma = 0$} sector.

%
%
\section{Discussion}
%
%
Interestingly, very similar results are found when studying the competition between double semion and  toric code phases \cite{Morampudi14}. This striking similarity has two main origins. First, the fact that, once the map of $\sigma^\pm$ is fixed, i.e., for a given set of $b_p^\sigma$, one can describe the present system by $\mathbb{Z}_2$-variables. Second, the model discussed in Ref.~\cite{Morampudi14} concerns two (Abelian) theories with different Frobenius-Schur indicator obeying $\mathbb{Z}_2$ fusion rules, and projectors onto the vacuum of these theories also obey Eq.~(\ref{eq:unit}).
  
We checked that the correspondence between both models also holds in the honeycomb lattice. Thus, one recovers a first-order transition between DIsing and D$\SU$ phases in the nontrivial low-energy sector at the self-dual points $|\tan \theta|=1$. However, the role played by frozen spins in two dimensions remain to be elucidated.\\

\acknowledgments

I would like to thank E. Ardonne, S. Dusuel, and V.~Lahtinen, S. H. Simon, and J. K. Slingerland for fruitful and insightful discussions.


\appendix

\section{Matrix elements of the projectors}
\label{app:Bp}
All information about the 16 theories obeying fusion rules given in Eqs.(\ref{eq:fusion1}) and (\ref{eq:fusion2}) can be found in Ref.~\cite{Rowell09}. In this Appendix, we simply give all ingredients to build the various projectors discussed in the main text. 

For Ising and $\SU$ theories, the modular $S$-matrix, in the basis $\{1,\sigma,\psi\}$ is given by: 
%
%
\begin{eqnarray}
\label{eq:deg_Fibo}
S &=&\frac{1}{D}\left(
\begin{array}{c c c}
1& \sqrt{2} & 1 \\
\sqrt{2} & 0 & -\sqrt{2} \\
1 & -\sqrt{2}& 1
\end{array}
\right),
\end{eqnarray} 
%
%
where \mbox{$D= \sqrt{\sum_i d_i^2}=2$} is the total quantum dimension, $(d_1,d_\sigma,d_\psi)=(1,\sqrt{2},1)$ being the quantum dimensions of particles  $1,\sigma$, and $\psi$, respectively. Frobenius-Schur indicators are  $\varkappa_1=\varkappa_\psi=+1$ for both theories whereas $\varkappa_\sigma=+1$ for Ising and $\varkappa_\sigma=-1$ for $\SU$.

The key ingredient to build the string-net Hamiltonian is the set of $F$-symbols \cite{Levin05,Gils09_1}. These $F$-symbols are defined as follows:
%
%
\begin{align}
\begin{array}{c}\includegraphics[width=1cm]{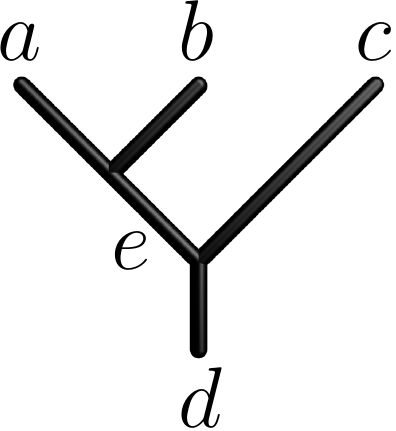}\end{array}
= 
\sum_f \left[F^{abc}_d \right] _{ef} \begin{array}{c}\includegraphics[width=1cm]{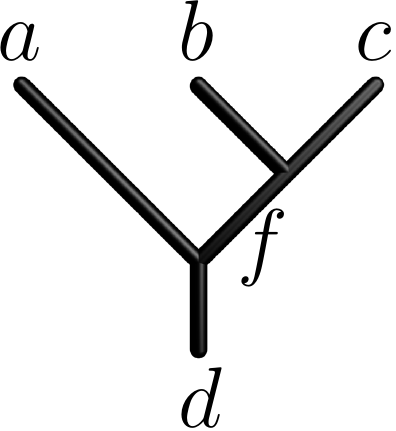}\end{array}.
\label{eq:Fmove}
\end{align}
%
%

In the basis $\{1,\psi\}$, one has \cite{Rowell09}

%
%
\begin{eqnarray}
F_\sigma^{\sigma \sigma \sigma} =\varkappa_\sigma \frac{1}{\sqrt{2}}\left(
\begin{array}{c c c}
1 & 1 \\
1 & -1\\
\end{array}
\right), \:  F_\sigma^{\psi \sigma \psi}=F_\psi^{\sigma \psi \sigma}=-1.\qquad
\end{eqnarray} 
%
%
Other $F$-symbols are equal to 1 if fusion channels are allowed and 0 otherwise.

%
%
\begin{figure}[h]
\includegraphics[width=0.65\columnwidth]{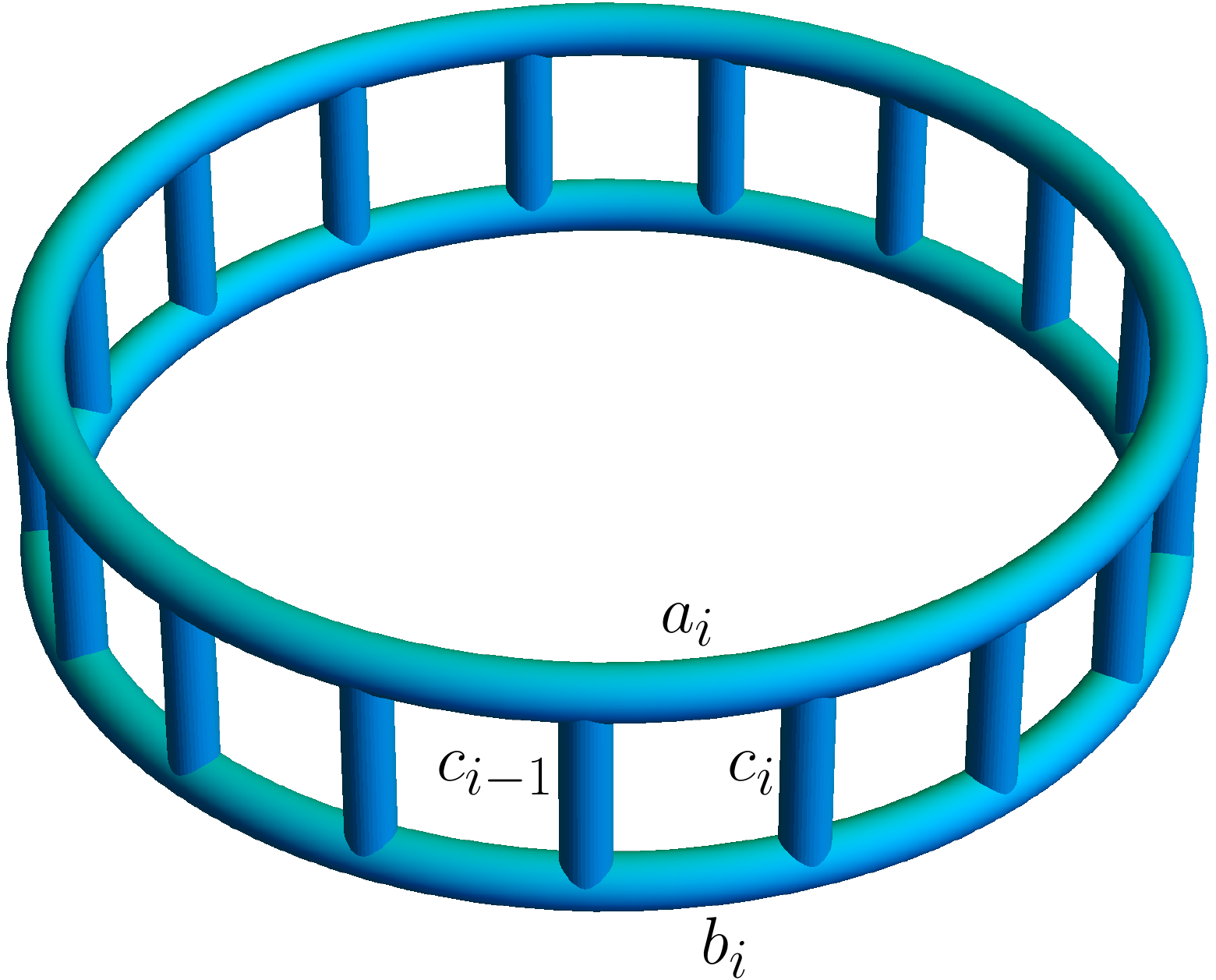}
\caption{Labeling of the links for a ladder with periodic boundary conditions.}
\label{fig:ladder_bare}
\end{figure}
%
%

\begin{widetext}

The projector onto the flux $\alpha^{\pm}=1^{\pm},\sigma^{\pm},\psi^{\pm}$ in the plaquette $i$  acts only on the plaquette links $(a_i,c_i,b_i,c_{i-1})$ (see Fig.~\ref{fig:ladder_bare}) as follows

%
%
\begin{equation}
\label{eq:projB}
B^{\alpha^{\pm}}_i |a_i,c_i,b_i,c_{i-1} \rangle = S_{1 \alpha}  \sum_\beta \varkappa_\beta S_{\alpha  \beta} 
\hspace{-5mm} \sum_{a'_i,c'_i,b'_i,c'_{i-1}}  \hspace{-2mm}
\Big[F_{b_{i}}^{b_{i-1} c'_{i-1} \beta}\Big]_{{c_{i-1}} b'_{i}} 
\Big[F_{c_{i}}^{b_{i+1} b'_{i} \beta}\Big]_{{b_{i}} c'_{i}} 
\Big[F_{a_{i}}^{a_{i+1} c'_{i} \beta}\Big]_{{c_{i}} a'_{i}} 
\Big[F_{c_{i-1}}^{a_{i-1} a'_{i} \beta}\Big]_{{a_{i}} c'_{i-1}} 
|a'_i, c'_i, b'_i, c'_{i-1} \rangle.  \nonumber 
\end{equation} 
%
%

Projectors onto the flux $\alpha^{\pm}=1^{\pm},\sigma^{\pm},\psi^{\pm}$ above and below the ladder act as follows
%
%
\begin{eqnarray}
\label{eq:projP}
P^{\alpha^{\pm}}_a |a,b,c \rangle&=& S_{1 \alpha} \sum_\beta \varkappa_\beta S_{\alpha  \beta} \sum_{a'} \prod_{i=1}^{N_{\rm p}}\Big[F_{a'_{i+1}}^{c_i a_i \beta}\Big]_{{a_{i+1}}{a'_i}} |a',b,c \rangle,  \quad \nonumber \\
P^{\alpha^{\pm}}_b |a,b,c \rangle&=& S_{1 \alpha}\sum_\beta \varkappa_\beta S_{\alpha  \beta} \sum_{b'} \prod_{i=1}^{N_{\rm p}} \Big[F_{b'_{i+1}}^{c_i b_i \beta}\Big]_{{b_{i+1}}{b'_i}} 
|a,b',c \rangle, \nonumber
\end{eqnarray} 
%
%
where $N_{\rm p}$ is the total number of plaquette and $|a,b,c \rangle$ is a state defined by the set of labels $\{a_i,b_i,c_i\}_{i=1,\cdots,N_{\rm p}}$. 
Depending on the superscript $+$ or $-$, $F$-symbols and Frobenius-Schur indicators of Ising or $\SU$ theory must be considered, respectively. 

Since  $S_{\sigma \sigma}=0$, one readily sees in these equations that only indices $\beta=1,\psi$ matter when $\alpha=\sigma$. Furthermore, reminding that all $F-$symbols of Ising and $\SU$ theories are the same, except for $F_{\sigma}^{\sigma \sigma \sigma}$\cite{Rowell09}, one readily gets \mbox{$P^{\sigma^{+}}_a=P^{\sigma^{-}}_a$, $P^{\sigma^{+}}_b=P^{\sigma^{-}}_b$}, and $B^{\sigma^{+}}_i=B^{\sigma^{-}}_i$.

\end{widetext}


\begin{thebibliography}{35}
\bibitem [{Con()}]{Configuration}%
  \BibitemOpen
  \href@noop {} {}\bibinfo {note} {Note that the four possible configurations
  of these two frozen spins have to be analyzed since the corresponding
  plaquettes can have two or no $\sigma$ legs. Configurations $|\! \! \uparrow
  \uparrow \rangle$, $|\! \! \downarrow \downarrow \rangle$ leads to the same
  spectrum as well as $|\! \! \uparrow \downarrow \rangle$ and $|\! \!
  \downarrow \uparrow \rangle$. When $N_{\rm p}$ is even, all of them are
  equivalent.}\BibitemShut {Stop}%
\end{thebibliography}

%

\end{document}